\documentstyle[psfig,12pt]{article}
\begin{document}
\title{Option Pricing Formulas  based on a non-Gaussian  Stock Price Model }
\author{Lisa Borland \\ Iris Financial Engineering and Systems\\456 Montgomery Street, Suite 800\\ San Francisco, CA 94104,  USA}
\maketitle

\begin{abstract}

Options are financial instruments that depend on 
the underlying stock. We explain their non-Gaussian fluctuations using the 
nonextensive thermodynamics   parameter $q$. A generalized form of the Black-Scholes (B-S) partial
differential equation, and  some closed-form solutions are obtained.
The standard B-S equation ($q=1$)  which is  used by economists to calculate option prices 
requires multiple values of the stock volatility (known as the  volatility smile).  Using
$q=1.5$ which well models the empirical distribution of returns, we  get a good description of option 
prices using a single volatility. 
 \end{abstract}

Although empirical stock price returns clearly do not follow 
the log-normal distribution,  many of the most famous results of mathematical finance are based 
on that distribution. For example, Black and Scholes (B-S) \cite{black&scholes} were able to derive the
  prices of options and other derivatives of the
 underlying stock based on such a model. An option is the right to buy or sell
the underlying stock at some set price (called the strike) at some time in the future. 
 While of great importance and widely used,
 such theoretical option prices do not quite
match the observed ones. In particular, the B-S model underestimates the prices
of options in situations when the stock price at the time of 
exercise is different from the strike. In order to match the observed market values, the B-S
model would need to use a different value of the volatility for each value of the strike. Such  
``implied volatilities'' of
options of various strike prices form a convex function known as the ``volatility smile''.

Indeed, attempts have been made to modify the B-S model in ways
that can correct for the smile effect (cf \cite{Hull} or more recently 
\cite{hyperbolic,bouchaud}). However, those approaches are often
very complicated or rather ad-hoc, and do not result in  managable closed form solutions, which is the
forte of the B-S approach. 
In this paper we do however succeed in developing  a theory of non-Gaussian option pricing which allows for 
closed form solutions for European options, which are such that can be exercised exclusively on a fixed day of expiration and not before (as is the case  for American options).

Our approach uses  stochastic processes with statistical feedback
\cite{pdependent} 
as a model for stock prices. Such processes were recently 
developed within the Tsallis generalized thermostatistics \cite{Tsallis}.
The driving noise  can be interpreted as 
a generalized Wiener process governed by a  Tsallis distribution of entropic index $q$.
In the limit $q \rightarrow 1$ the standard model is recovered. 
For $q \approx 1.5$, this model closely fits the
empirically observed distribution for many financial time series, such as stock prices \cite{Osorioetal}
 (Figure 1),
SP500 index,\cite{Osorioetal})\cite{Michael&Johnson},  FX rates, etc. 
This is  consistent with a cumulutive distribution having power tails of index 3 \cite{gopik}.
We  derive closed form option
pricing formulas, reproducing  prices which, relative to the standard B-S
model, exhibit volatility smiles very close to those observed empirically (Figure 4).  
Note that $q=1.5$ well models hydrodynamic
turbulence on small scales \cite{beck}, reinforcing notions of a possible 
analogy between these two systems.

The standard model for stock prices is 
$
S = S_0 e^{Y(t)}
$
where $Y(t) = \ln(S(t+ t_0)/\ln S(t_0) $ follows
\begin{equation}
\label{eq:lanstand}
d Y = \mu dt + \sigma d \omega
\end{equation}
The drift $\mu$ is the mean rate of return and $\sigma^2$ is the variance of the
 stock logarithmic return. The noise 
$\omega$ is a Brownian motion defined with respect to a probability measure $F$. It is a  Wiener process and 
satisfies $
E^F[d\omega(t)d\omega(t')] = dtdt'\delta(t-t')
$
where the notation  $E^F[]$ means the expectation value with respect to the measure $F$.
This model yields a Gaussian distribution for $Y$ resulting in
a log-normal distribution for  $S$.
Within this framework,  Black and Scholes were able
 to establish 
a pricing model  to obtain the fair value of options on the underlying stock $S$.
 
In this paper we assume that 
the log returns $Y(t) = \ln S(t + t_0) /\ln S(t_0)$ follow 
\begin{equation}
\label{eq:langen}
d Y = \mu dt + \sigma d \Omega
\end{equation}
with respect to the timescale  $t$.
Here  $\Omega$ evolves
according to  the statistical feedback process\cite{pdependent}
\begin{equation}
\label{eq:nomega}
d \Omega =  P(\Omega)^{\frac{1-q }{2}} d \omega
\end{equation}
The probability distribution $P$ satisfies the nonlinear Fokker-Planck
equation
\begin{equation}
\label{eq:nlfpcond}
\frac{ \partial}{\partial t} P(\Omega, t \mid \Omega ', t') = \frac{1}{2} \frac{ \partial }{\partial \Omega ^2} P^{2-q}(\Omega, t \mid \Omega ', t')
 \end{equation}
Explicit solutions for $P$ are given by Tsallis distributions
\cite{Tsallis&Bukman}
 \begin{equation}
\label{eq:ptsalliscond}
P_q(\Omega, t \mid \Omega(0), 0) = \frac{1}{Z(t)} \{ 1 - \beta(t) (1-q)[\Omega(t) - \Omega(0)] ^2 \}^{\frac{1}{1-q}}
\end{equation}
Choosing
$
\beta(t) = c^{\frac{1-q}{3-q}}((2-q) (3-q) t)^{-2/(3-q)} $ 
and 
$ Z(t)  =  ( (2-q) (3-q) c t )^{\frac{1}{3-q}} 
$
ensures that the initial condition $P_q = \delta(\Omega(t) - \Omega(0)) $
is satisfied.
The   $q$-dependent constant $c$ is given by
$
c  =  \beta Z^2 $ with 
$Z  = \int_{- \infty}^{\infty} ( 1 - (1-q)\beta \Omega^2)^{\frac{1}{1-q}} d \Omega
$
for any $\beta$. 
With  $\Omega(0) = 0$, we obtain a generalized Wiener 
process, distributed according to a zero-mean Tsallis distribution
In the limit $q \rightarrow 1$ the standard theory Eq(\ref{eq:lanstand}) is recovered, 
and $P_q$ becomes a Gaussian.
We are concerned with the range $1 \le q < 5/3$ in which positive 
tails and finite variances are found \cite{Levy}.
The distribution  for $\ln S$ becomes
\begin{equation}
\label{eq:ptsallisnonzero}
P_q(\ln S(t+t_0),t+t_0 \mid \ln S(t_0), t_0) = \frac{1}{Z(t)} \{1 - \tilde{\beta}(t) (1-q)[\ln \frac{S(t+t_0)}{S(t_0)} - \mu t ] ^2 \}^{\frac{1}{1-q}}
\end{equation}
with  $\tilde{\beta}   = \beta(t)/ \sigma ^2 $.
This implies that log-returns $\ln [S(t +t_0)/S(t_0)]$ 
over the timescale  $ t$ follow a Tsallis distribution, consistent with empirical evidence for several markets, e.g. S\&P500 (Figure 1 \cite{Osorioetal}) \cite{Michael&Johnson}, with $q \approx 1.5$.

 Our  model exhibits a 
  feedback from the macroscopic level characterised
 by $P$, to the microscopic level characterised by  $\Omega$.
We can imagine that this is really due to the 
interactions of many individual traders whose actions all  contribute to
shocks to  the stock price which  keep it in equilibrium. Their collective behaviour 
yields a nonhomogenous 
 reaction to returns: rare events (i.e.
 extreme returns) will be accompanied by large reactions, and will tend to be followed by 
large returns in either direction.

Using Ito calculus \cite{Gardiner, Risken}, the equation for $S$ follows from Eq(\ref{eq:langen}) as 
\begin{equation}
\label{eq:langenS}
d S = \tilde{\mu} Sdt  + \sigma S d \Omega
\end{equation}
where $
\tilde{\mu} = \mu + \frac{\sigma^2}{2} P_q^{1-q}. $
The term $\frac{\sigma^2}{2} P_q^{1-q}$  is  the   noise induced drift.
Remember that $P_q$ is a function of  $\Omega$ with 
\begin{equation}
\label{eq:omegaands}
\Omega(t) = \frac{{\ln S(t)}/{\ln S(0)} - \mu t}{\sigma}
\end{equation}
(with $t_0=0$ for simplicity.)
As in the standard
case (cf \cite{Hull}), the noise term driving
 $S$ is the same as that driving the price $f(S)$ of a  derivative of the underlying stock. 
It should be possible to invest one's wealth in a portfolio of shares and 
derivatives such that the noise terms cancel each other, yielding 
a risk-free portfolio,  the return on which must be the
risk-free rate $r$. This  results in a generalized B-S PDE
\begin{equation}
\label{eq:bsgen}
\frac{\partial f}{\partial t}  + rS \frac{\partial f}{\partial S}  + \frac{1}{2} \frac{\partial^2f}{\partial S^2} \sigma^2 
S^2 P_q^{1-q} 
= rf  
\end{equation}
where $P_q( \Omega(t))$ evolves according to Eq(\ref{eq:nlfpcond}).
For $q \rightarrow 1$ the standard case is recovered.
This PDE  depends explicitly only on the risk-free rate and the variance,
not on $\mu$, but it does depend implicitly on $\mu$ through
$P_q(\Omega)$, with $\Omega$ given by Eq(\ref{eq:omegaands}).
Therefore, to be consistent with
 risk-free pricing theory, we should first transform our original stochastic 
equation for $S$
 into a  martingale before we apply the above analysis. This will
 not affect our results 
other than that $\tilde{\mu}$ will be replaced by the risk-free rate $r$,
ultimately eliminating the dependency on $\mu$. 
We now show how  this is done.

The discounted stock price $G = e^{-rt}S$ follows 
$
dG = (\tilde{\mu} -r)Gdt + \sigma G d \Omega
$
where $d \Omega$ follows Eq(\ref{eq:nomega}).
For there to be no arbitrage opportunities, risk-free asset pricing theory requires
that this process be a martingale, which it is not due to the
 drift term
$(\tilde{\mu} - r)G dt$. One can however define an alternative driving noise $z$ associated
with an equivalent probability  measure $Q$ so that, with respect to the new noise measure, 
the discounted stock price has zero drift and is thereby a martingale. Explicitly,
\begin{equation}
\label{eq:dGomega}
dG  =  (\tilde{\mu} -r)Gdt + \sigma G P^{\frac{1-q}{2}} d \omega
\end{equation}
Here, $P$ is a non-vanishing bounded function of $\Omega$. With respect to the initial 
noise $\omega$, $\Omega$ relates to $S$ via Eq(\ref{eq:omegaands}). That is why 
for all means and purposes, $P$ in Eq(\ref{eq:dGomega}) is simply a function of $S$ (or $G$), 
and the stochastic process can be seen as a standard state-dependent Brownian one. As a consequence,
both the Girsanov theorem (which specifies the conditions under which we can 
transform from the measure $F$ to $Q$)
 and the Radon-Nikodym theorem (which relates the measure $F$ to $Q$) are valid, and we can formulate equivalent martingale measures much as in the standard case 
\cite{Oksendal,Marek&Marek,Shreve}.
We rewrite Eq(\ref{eq:dGomega}) as
\begin{eqnarray}
\label{eq:dGmart}
dG & = & \sigma G P^{\frac{1-q}{2}} d z
\end{eqnarray}
where the  new driving noise term $z$ is related to $\omega$ through 
\begin{equation}
\label{eq:dz}
dz = \frac{(\tilde{\mu} -r)}{\sigma P_q^{\frac{1-q}{2}}} dt + d \omega 
\end{equation}
With respect to $z$, we thus obtain
$
dG  =  \sigma G d\Omega
$
with
$
d \Omega = P_q^{\frac{1-q}{2}} dz
$
which  is non other  than a zero-mean Tsallis distributed generalized Wiener process,
completely analogous to the  one defined in Eq(\ref{eq:nomega}).
Transforming back to $S$ yields
$
dS  = r dt + \sigma S d\Omega.
$
Compared with Eq(\ref{eq:langenS}), the 
rate of return $\tilde{\mu}$ has been replaced with the risk-free rate $r$.
This recovers the same result as in the standard asset pricing theory.
Consequently, in the risk-free representation, Eq(\ref{eq:omegaands}) becomes
\begin{equation}
\Omega(t) = \frac{1}{\sigma}(\frac{\ln S(t)}{\ln S(0)} - r t + \frac{\sigma^2}{2}\int_{0}^t P_q^{1-q}( \Omega(s)) ds)
\end{equation} 
This eliminates the dependency on $\mu$ which we alluded to in the discussion
of Eq(\ref{eq:bsgen}). 
As discussed later on, by standardizing the  distributions $P_q(\Omega(s))$ we can 
explicitly solve for $\Omega(t)$ as a function of $S(t)$ and $r$.

Suppose that we have a European option  $C$ which depends on $S(t)$, whose 
price $f$  is given by its expectation value in a risk-free (martingale)
 world  as
$
f(C) = E^Q [e^{-rT} C]
$.
We assume the payoff on this option depends on the stock price at time $T$ so that
$
C = h(S(T))
$. After stochastic integration of Eq(\ref{eq:dGmart}) to obtain $S(T)$ we get
\begin{equation}
\label{eq:fprice1}
f = e^{-rT} E^Q\left[h \left( S(0) \exp \left( \int_0^T \sigma P_q^{\frac{1-q}{2}} dz_s + \int_0^T(r - \frac{\sigma^2}{2} P_q^{1-q}) ds \right) \right) \right]
\end{equation} 
The key point is that the
 random variable   
$
 \int_0^T P_q^{\frac{1-q}{2}} dz_s  =  \int_0^T d \Omega(s) =  \Omega(T) 
$
follows the Tsallis distribution Eq(\ref{eq:ptsalliscond}). 
This gives
\begin{eqnarray}
\label{eq:generalformula}
f &=& \frac{e^{-rT}}{Z(T)} \int_R h\left[ S(0)\exp(\sigma \Omega(T) +rT - \frac{\sigma^2}{2} \alpha T^{\frac{2}{3-q}} 
 + (1-q) \alpha T^{\frac{2}{3-q}} \frac{\beta(T)}{2} \sigma^2 \Omega^2(T) )  \right]
\nonumber \\ & & (1-{\beta}(T)(1-q)\Omega(T)^2)^{\frac{1}{1-q}} d\Omega_T
\end{eqnarray}
with $
\alpha =  \frac{1}{2}(3-q) ((2-q)(3-q))c)^{\frac{q-1}{3-q}}.
$
We have utilized the fact that each of
the distributions $P(\Omega(s))$ occuring in the latter term of  Eq(\ref{eq:fprice1}) can be mapped  onto the 
distribution of $\Omega(T)$  at time $T$ 
via the appropriate variable transformations
$
\Omega(s)  = \sqrt{ {\beta(T)}/{\beta(s)} } \Omega(T)
$.
A major difference to the standard case is the $\Omega^2(T)$-term  
which is a result of the noise induced drift. 
With $q= 1$, the standard  option price is recovered \cite{Oksendal}.

Eq(\ref{eq:generalformula}) is valid for an arbitrary payoff $h$. We shall
evaluate it explicitly for 
a European call option,  which gives the holder  the right to 
buy  the  stock $S$ at the strike price $K$, on the day of expiration 
$T$. The payoff is
$
C = \max [ S(T) - K, 0]
$.
Only if  $S(T) > K$ will the option have value at expiration $T$ 
(it will be in-the-money).
The price $c$ of such 
an option becomes
\begin{eqnarray}
\label{eq:optionprice}
c & = & E^Q [e^{-rT} C]
  =  E^Q [ e^{-rT} S(T)]_D - E^Q [ e^{-rT} K]_D 
 =  J_1 - J_2 \label{eq:cprice}
\end{eqnarray}
where the subscript $D$ stands for the set $\{ S(T) > K \}$.
This condition is met if 
$
-\frac{\sigma^2}{2} \alpha T^{\frac{2}{3-q}} +
(1-q)\alpha T^{\frac{2}{3-q}} \frac{\beta(T)}{2} \sigma^2 \Omega^2 + \sigma \Omega + rT > \ln {K}/{S(0)},
$
 which is satisfied for $\Omega$ between the two roots $s_1$ and $s_2$ of the
corresponding quadratic equation.
This is a very different situation from the standard case, where the inequality
is linear and the condition $S(T) > K$ is satisfied for all values of the
random variable greater than a threshold. In our case, due to the noise induced
 drift, values of $S(T) $ in the risk-neutral world are not monotonically 
increasing as a function of the noise.  As $q \rightarrow 1$, the larger 
root $s_2$ goes toward $\infty$, recovering the standard case. But as $q$ increases, the tails of the noise distribution get larger, as does the noise induced drift which tends to pull the system back. 
As a result we obtain
\begin{eqnarray}
J_1
\label{eq:j1} 
&  = & S(0) \frac{1}{Z(T)} \int_{s_1}^{s_2} \exp( \sigma \Omega -
\frac{\sigma^2}{2} \alpha T^{\frac{2}{3-q}} 
 - (1-q) \alpha T^{\frac{2}{3-q}} \frac{\beta(T)}{2} \sigma^2 \Omega^2 ) \nonumber \\
& & (1 - (1-q)  \beta(T) \Omega^2)^{\frac{1}{1-q}} d\Omega  \\
\label{eq:j2}
J_2 & = & e^{-rT} K \frac{1}{Z(T)} \int_{s_1}^{s_2} (1 - (1-q) {\beta}(T) \Omega^2)^{\frac{1}{1-q}} d\Omega 
\end{eqnarray}

The equation Eq(\ref{eq:cprice}) with Eq(\ref{eq:j1}) and Eq(\ref{eq:j2}) constititutes
a closed form expression for the price of a European call. 
We calculated option prices for different values of the
 index $q$, and studied their properties as a function of the relevant variables such as 
the current stock price $S(0)$, the strike price $K$, time to expiration $T$, the risk free
 rate $r$ and $\sigma$. The results obtained by our closed form pricing formula  were 
confirmed both  by implicitly solving the generalized B-S PDE Eq(\ref{eq:bsgen}) 
and via Monte Carlo simulations of Eq(\ref{eq:dGmart}). 
Note that American option prices can  be solved numerically via Eq(\ref{eq:bsgen}).

We compare results of the standard model
($q = 1$) with those obtained for $q= 1.5$, which 
fits well to  real stock returns. Figure 2 shows the difference in call price. 
In Figure 3, the B-S implied volatilities (which make the $q=1$ model match the $q=1.5$ results) are plotted as a function of $K$.
The assymetric smile shape, which is more pronounced for shorter times, reproduces well-known systematic features of the
 ``volatility smile'' that appears when using the standard $q=1$ model to
 price real options. 
In Figure 4,  the volatility smiles for actual
 traded options on BP and S\&P 500 futures is shown  together with those resulting from our model using $q=1.5$.
These results are encouraging, and we are currently studying  a larger sample of 
options data. Empirical work is required  to see  if arbitrage opportunities can be uncovered that do not
appear when the standard model is used.  Another potential  
application will be with respect to option replication and hedging. 

Acknowledgements: Fruitful discussions with Roberto Osorio and Jeremy Evnine are gratefully
acknowledged.

\begin{figure}[t]
\psfig{file=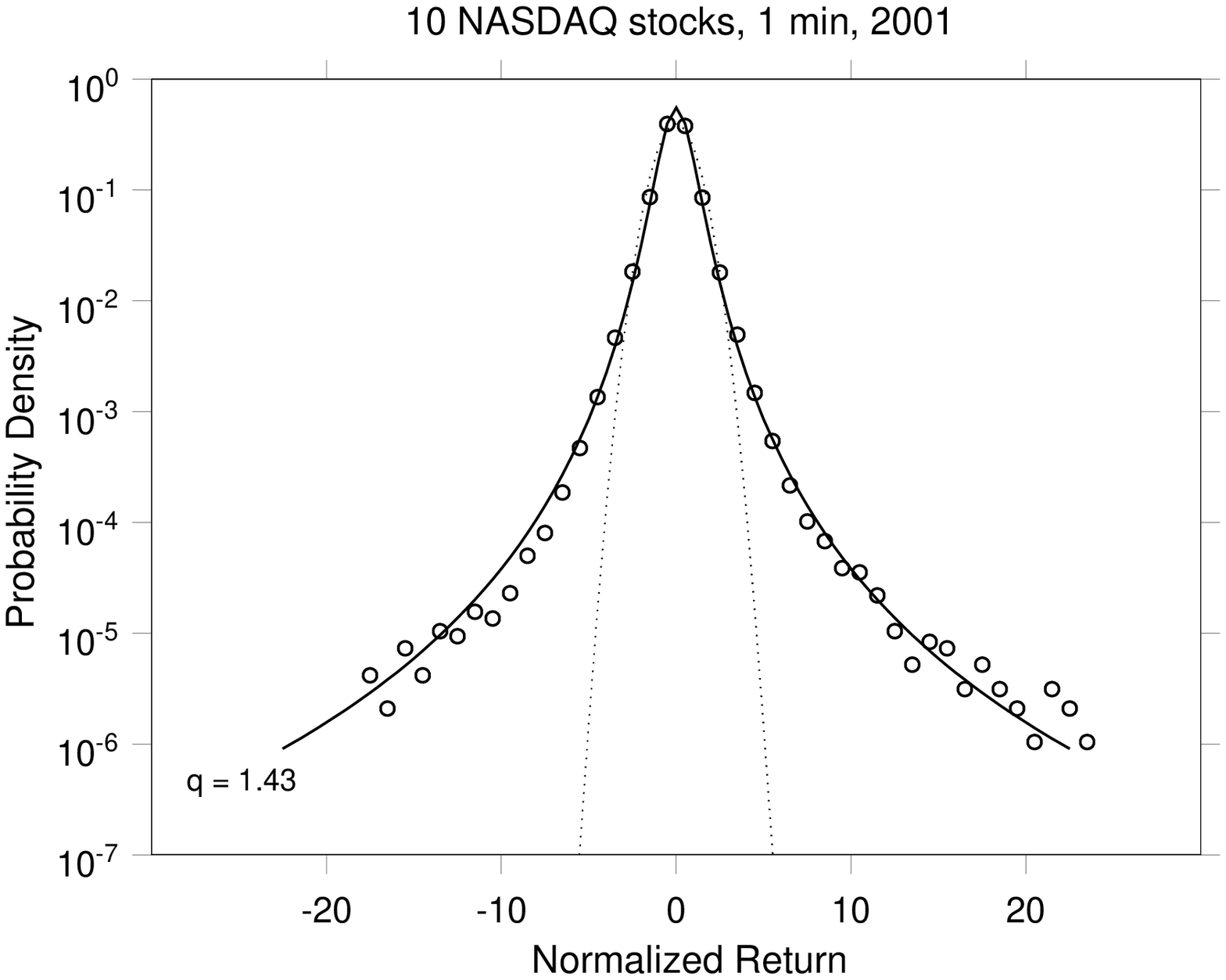,width= 5.5in}
\caption{\footnotesize
Distributions of log returns over 1 minute intervals for 10 high-volume
 stocks, normalized by the sample standard deviation. Also shown is the Tsallis distribution of index $q = 1.43$ (solid line) which provides a good fit to the data (Figure kindly provided by R. Osorio). 
}
\end{figure}

\begin{figure}[t]
\psfig{file=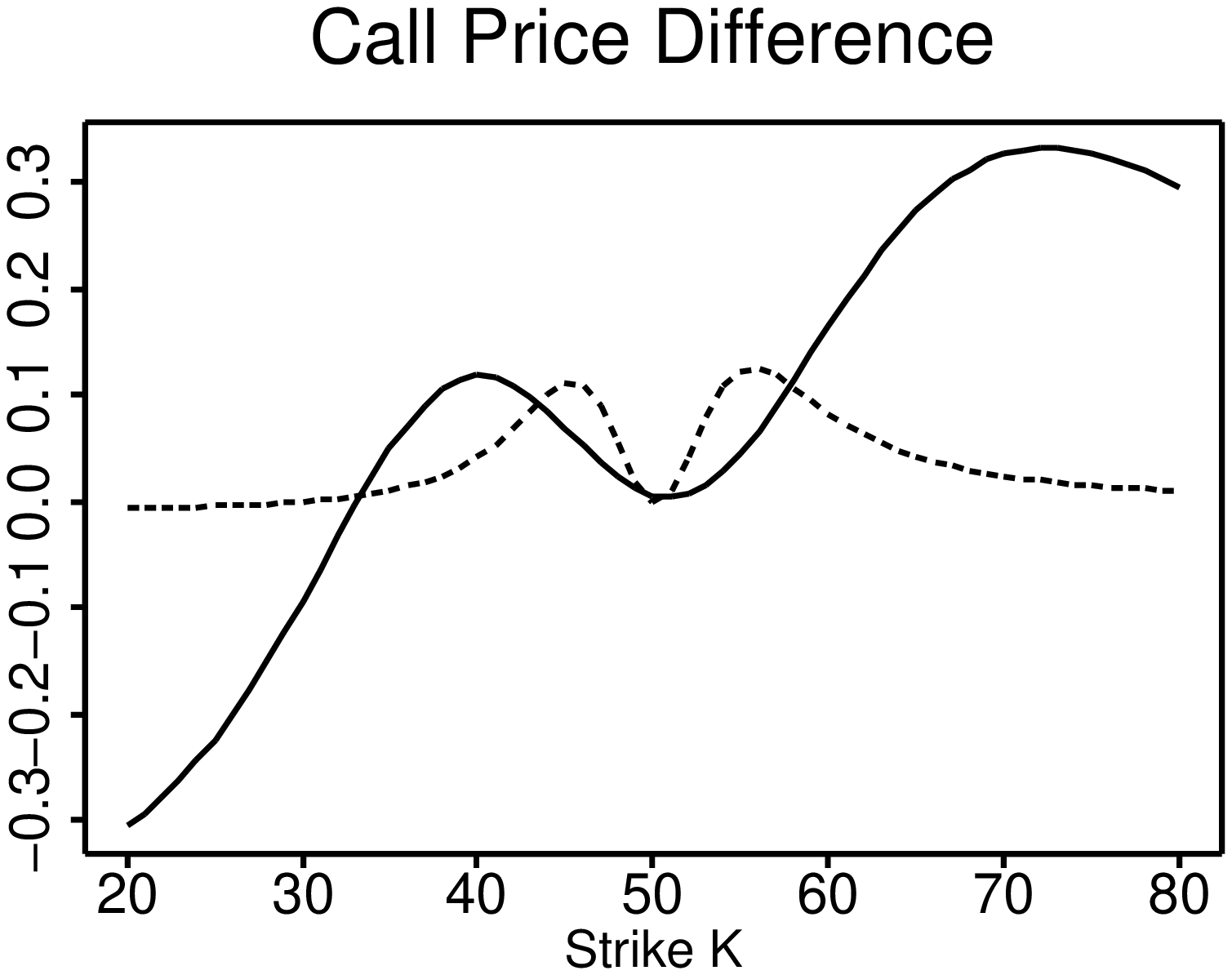,width=4.5in}
\caption{\footnotesize
Calibrated so that the options are priced equally for $S(0)=K$, the difference
between the $q=1.5$ model and the standard B-S model is shown, for  $S(0) = \$ 50$ and  $r= 0.06$. Solid line: $T=0.6$  with $\sigma = .3$
for $q=1$ and  $\sigma = .297$  for $q=1.5$. Dashed line: $T=0.05$ 
with $\sigma = .3$ for $q=1$ and $\sigma = .41$ for  $q=1.5$.  Times are 
expressed in years, $r$ and $\sigma$ are in annual  units.}
\end{figure}

\begin{figure}[t]
\psfig{file=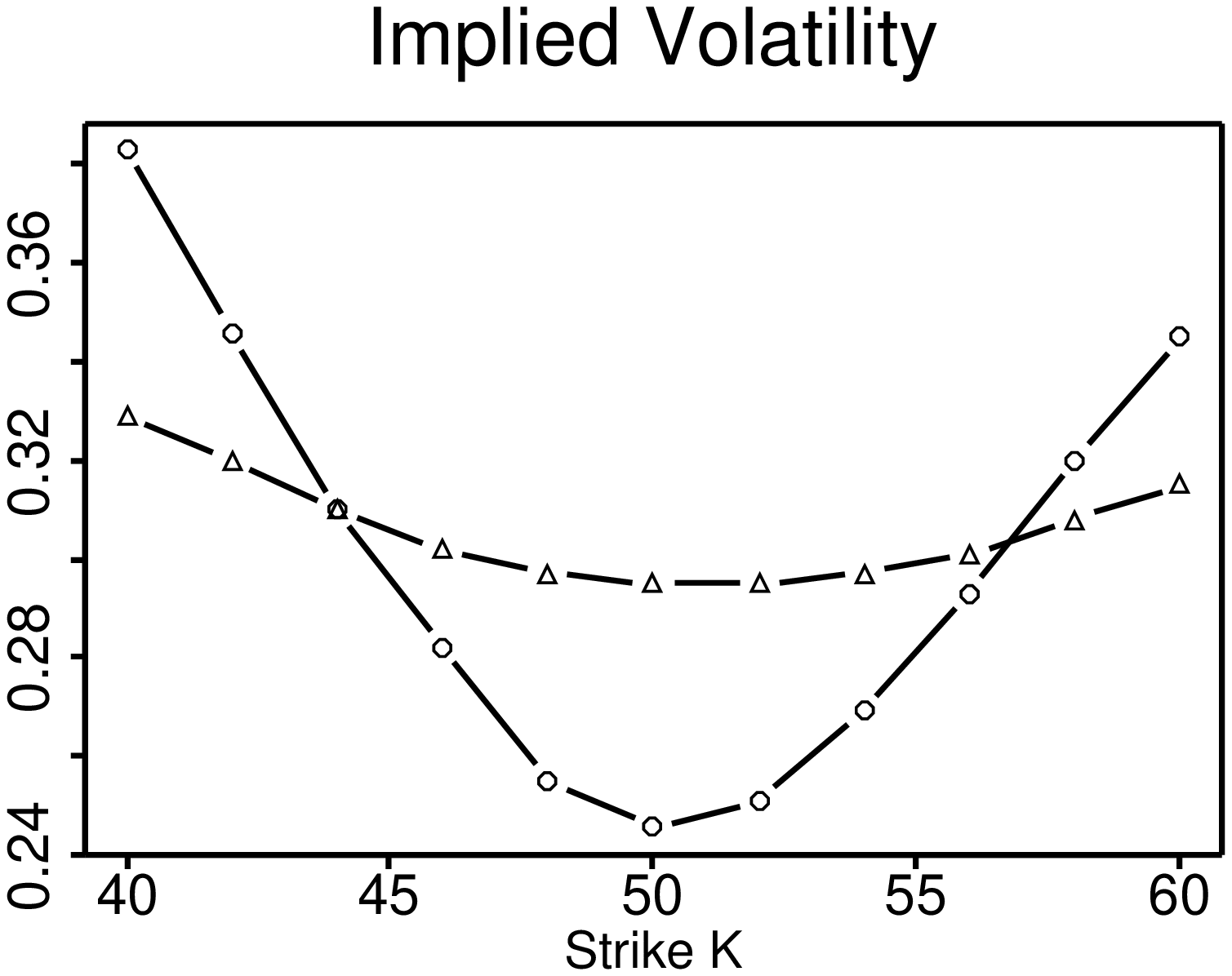,width=4.5in}
\caption{\footnotesize
Using the $q=1.5$ model (with $\sigma =.3$, $S(0) = 50 \$$ and $r=.06$) to generate call option prices, one can back out
the volatilities implied by a standard $q=1$ B-S model.
 $T=0.1$ (circles), $T= 0.4$ (triangles).
}
\end{figure}

\begin{figure}[t]
\psfig{file=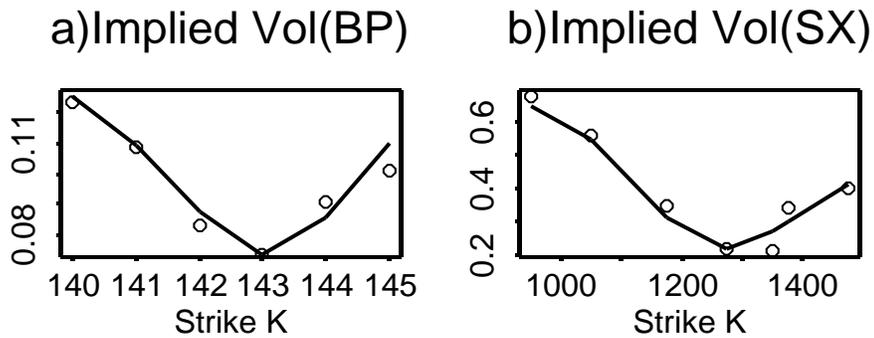,width=4.5in}
\caption{\footnotesize
A comparison of option prices from a $q=1.5$ model and traded prices is given
 by a comparison of volatility smiles.
 a)Implied vols. for options on British Pound futures (last trading date Dec 7, 2001) \cite{bpfutures}
vs. strike (for $S(0)= 143$, $r=.065$, $T=.0055$ (2 days)) (symbols); implied
vols needed for a  $q=1$ B-S model to match 
prices from a $q=1.5$ model using  $\sigma=.1445$ across all strikes (line).b) Implied vols for S\&P500 futures (SX June, last trading date June 15, 2001, $S(0) = 1275, r=.065, T=.027$ (10 days)) (symbols); implied vols. from $q=1.5$ model with $\sigma =.3295$ (line). 
}
\end{figure}

\end{document}